\documentclass[sigplan,screen]{acmart}
\usepackage{booktabs}
\usepackage{subcaption}
\usepackage{graphicx}
\usepackage[final]{listings}
\usepackage{hyperref}
\usepackage{amsmath}
\usepackage{amsfonts}
\usepackage{upquote}
\usepackage[tableposition=top]{caption}
\usepackage{enumitem}
\usepackage{multicol}
\usepackage{circledsteps}
\usepackage[utf8]{inputenc}
\usepackage[T1]{fontenc}
\usepackage{lineno}
\usepackage{ifmtarg}
\usepackage{relsize}
\usepackage{calc}
\usepackage{microtype}

\usepackage{afterpage}

\ttfamily

\makeatletter
\DeclareFontFamily{T1}{lmtt}{\hyphenchar \font\m@ne}
\DeclareFontShape{T1}{lmtt}{l}{n}{<->ec-lmtl10}{}
\DeclareFontShape{T1}{lmtt}{l}{it}{<->sub*lmtt/l/sl}{}
\DeclareFontShape{T1}{lmtt}{l}{sl}{<->ec-lmtlo10}{}
\makeatother
\newcommand{\lmttfamily}{\fontfamily{lmtt}\fontseries{l}\selectfont}

\newcommand{\csharp}{C\nolinebreak\hspace{-.05em}\raisebox{.6ex}{\scriptsize\textbf\#}}
\newcommand{\fsharp}{F\nolinebreak\hspace{-.05em}\raisebox{.6ex}{\scriptsize\textbf\#}}

\definecolor{dark-green}{rgb}{0,0.5,0}
\definecolor{comment-red}{rgb}{0.8,0,0}
\definecolor{host}{rgb}{1,0,0}
\definecolor{exp}{rgb}{0,0,1}

    \newcommand{\note}[2]{}
    \newcommand{\TODO}[1]{}
    
\newcommand{\hostTypeTAG}{TAG}  
\newcommand{\hostTag}[1]{{Const} {#1}}
\makeatletter
\newcommand{\embeddedTag}[1]{$\langle {\mathtt{TRACE}}$\@ifnotmtarg{#1}{~#1}$\rangle$}
\makeatother

\lstdefinestyle{haskell}{%
    language=Haskell,
    upquote=true,
    deletekeywords={case,class,data,newtype,default,deriving,do,in,instance,let,of,type,where,if,then,else},
    morekeywords={[2]class,data,newtype,default,deriving,family,instance,type,where,pattern},
    morekeywords={[3]in,let,case,of,do,if,then,else},
    literate=
        {\\\\}{{\char`\\\char`\\}}1
        {>->}{>->}3
        {>>=}{>>=}3
        {->}{{$\rightarrow$}}2
        {>=}{{$\geq$}}2
        {<-}{{$\leftarrow$}}2
        {<=}{{$\leq$}}2
        {=>}{{$\Rightarrow$}}2
        {|}{{$\mid$}}1
        {~}{{$\sim$}}1
        {forall}{{$\forall$}}1
        {exists}{{$\exists$}}1
        {...}{{$\cdots$}}3
}

\lstdefinestyle{llvm}{
    language=LLVM,
    escapechar={|},
    keywordstyle=[2]\bf,
}

\lstdefinestyle{inline}{
    basicstyle=\small\ttfamily,
    keywordstyle=[1],
    keywordstyle=[2],
    keywordstyle=[3],
    keywordstyle=[4],
}

\lstdefinestyle{footnote}{
    basicstyle=\footnotesize\ttfamily,
    keywordstyle=[1],
    keywordstyle=[2],
    keywordstyle=[3],
    keywordstyle=[4],
}

\newcommand{\makeatcode}{\lstMakeShortInline[style=inline]@}
\newcommand{\makeatchar}{\lstDeleteShortInline@}

\lstnewenvironment{haskell}
  {\lstset{style=haskell}}
  {}

\setcopyright{acmcopyright}
\acmPrice{15.00}
\acmDOI{10.1145/3546189.3549917}
\acmYear{2022}
\copyrightyear{2022}
\acmSubmissionID{icfpws22haskellmain-p33-p}
\acmISBN{978-1-4503-9438-3/22/09}
\acmConference[Haskell '22]{Proceedings of the 15th ACM SIGPLAN International Haskell Symposium}{September 15--16, 2022}{Ljubljana, Slovenia}
\acmBooktitle{Proceedings of the 15th ACM SIGPLAN International Haskell Symposium (Haskell '22), September 15--16, 2022, Ljubljana, Slovenia}
\begin{document}

\title{Embedded Pattern Matching}

\author{Trevor L. McDonell}
\email{t.l.mcdonell@uu.nl}
\orcid{0000-0001-7806-9751}
\affiliation{%
  \institution{Utrecht University}
  \country{Netherlands}
}
\author{Joshua D. Meredith}
\email{josh.meredith@iohk.io}
\orcid{0000-0003-0797-4329}
\affiliation{%
  \institution{IOG}
  \country{Australia}
}
\author{Gabriele Keller}
\email{g.k.keller@uu.nl}
\orcid{0000-0003-1442-5387}
\affiliation{%
  \institution{Utrecht University}
  \country{Netherlands}
}



%
%
%
%
%
%
%
%
%
%
\begin{abstract}
  Haskell is a popular choice for hosting deeply embedded languages. A recurring
  challenge for these embeddings is how to seamlessly integrate user defined
  algebraic data types. In particular, one important, convenient, and expressive
  feature for creating and inspecting data---pattern matching---is not directly
  available on embedded terms.
  We present a novel technique, \emph{embedded pattern
  matching}, which enables a natural and user friendly embedding
  of user defined algebraic data types into the embedded language,
  and allows programmers to pattern match on terms in the embedded language
  in much the same way they would in the host language.
\end{abstract}

\ccsdesc[500]{Software and its engineering~Functional languages}
\ccsdesc[500]{Software and its engineering~Domain specific languages}
\ccsdesc[500]{Software and its engineering~Data types and structures}

\keywords{Haskell, pattern matching, algebraic data types, embedded languages}
\maketitle
\makeatcode

%
%

\section{Introduction}
\label{sec:introduction}

Algebraic data types have proven to be a powerful, convenient, and expressive
way of creating and organising data, and have become a defining
characteristic of functional programming languages in particular.
%
%
Algebraic data types and pattern matching are highly suited to
defining abstract syntax trees and the operations on them. This
enables a popular and time-saving way to define and implement a
[domain-specific] programming language: by \emph{embedding} it in
another one. Embedding means to represent the terms and values of the
\emph{embedded language} as terms and values in the \emph{host
  language}, so that the former can be interpreted in the
latter~\cite{Reynolds:1972}. While a \emph{shallow} embedding executes the
operations of the embedded language directly, a \emph{deep} embedding instead builds an
abstract syntax tree (AST) that represents the embedded program.
For example, the abstract syntax tree for the untyped lambda calculus can be defined as:
\begin{haskell}
data Name   = Name String
data Lambda = Var Name
            | App Lambda Lambda
            | Lam Name Lambda
\end{haskell}
%
%
Writing an evaluation function for terms in this language is a straightforward
exercise, and more complex transformations such as optimisation and code
generation are also possible.

The most light-weight (and probably most common) approach to obtaining such
an AST is what we call a \emph{combinator-based deep embedding}, where the
language terms more or less directly call the constructors of the
AST~\cite{Feldspar,Chakravarty:2011,Gill:2009wy,Rompf:2010,Elliott:1997b}. This allows the
language implementor to use most of the host language compiler's infrastructure.
Depending on the expressivity of the host language type system, many of the
sanity checks for the embedded language can also be delegated to the host
language type checker. Using overloading, this type of deep embedding can be
made almost completely transparent to the user, and gives the illusion of
working in the host language directly.


Other popular approaches, which construct the AST by using
template- or meta-programming~\cite{Torlak:2013,Najd:2016},
or compiler plug-ins~\cite{plutus,Young:2021} require more
implementation effort, but in exchange offer more flexibility compared to the
combinator-based approach as they reuse less functionality of the host
language (for example, by implementing their own parser and type checker) and may
manipulate the AST of the host language itself. One drawback of
combinator-based deep embeddings, on the other hand, is that user-defined algebraic data types and
pattern matching---the very features which make implementing the embedding
itself so convenient---are themselves difficult to integrate into the
embedding. Consequently, these important language features are not well supported, if at all.

In this paper we present a novel technique to support user-defined algebraic
data types in combinator-based deep embeddings.
While some embedded languages support product types to some extent, to the best
of our knowledge there are no combinator-based deeply embedded languages which support sum or
recursive data types, nor pattern matching on embedded values of algebraic data
type.
Our solution to the problem is based on two ideas. First, we
automatically map algebraic data types to a generic internal
representation which is amenable to inspection. Second, we develop a
technique so that the pattern matching mechanism of the \emph{host} language can be
used to construct the abstract syntax for case distinctions in the
\emph{embedded} language. These techniques are sufficiently flexible to handle arbitrarily
nested algebraic data types, including user-defined ones.
Both the mapping to the internal representation and the pattern matching in the
embedded language happen almost transparently to the user, so that the code
they have to write to create, inspect, and manipulate values of algebraic data
type in the embedded language closely resembles that of the host
language. For example, we can write an \emph{embedded} program that converts
\emph{embedded} lambda terms into SKI combinator form as:
\begin{haskell}
toSKI :: %\textcolor{exp}{Exp Lambda}% -> %\textcolor{exp}{Exp SKI}%
toSKI = match \case
  Var_ v   -> ...
  App_ f x -> ...
  Lam_ v x -> ...
\end{haskell}
Our technique is not merely one of convenience: it also exposes
optimisation opportunities which are obfuscated if the user has to resort to
workarounds in the absence of these features.
%
In this paper we present:
\begin{itemize}[leftmargin=*]
  \item a mapping for arbitrary user-defined algebraic
    data types to a generic internal representation in the abstract syntax tree
    of an embedded program (\S\ref{sec:expression});
  \item embedded pattern matching, a user-friendly technique to lift
    pattern matching from the host language into the embedded language
    (\S\ref{sec:implementation}); and
  \item a case study of these techniques implemented for an existing
    real-world embedded language, and discuss the impact of these
    changes both in terms of the user experience and efficiency of the
    generated code (\S\ref{sec:evaluation}).
\end{itemize}

We demonstrate our technique using a core expression language
embedded in Haskell, but our technique is applicable to other host languages.
Section~\ref{sec:embedded_languages} gives an overview of deep embeddings,
and the problem algebraic data types pose in this setting.
Section~\ref{sec:pattern_matching} briefly covers the syntax and semantics
of pattern matching in Haskell. 
Section~\ref{sec:related} discusses related work.

The source code for the implementation described in this paper is
available at 
\url{https://github.com/tmcdonell/embedded-pattern-matching}.

\section{Deeply Embedded Languages}
\label{sec:embedded_languages}

The power of deeply embedded languages is that the implementer has
complete control over the evaluation of the program and can support
multiple interpretations of it. The disadvantage is that realising the
embedding in a user-friendly way is much more challenging than in the
shallow case, because we are essentially hiding from the user that
they are working on syntax trees representing the computation of
values, rather than computing on values directly.

Ideally, the user of an embedded language would be able to use all of
the features of the host language in the embedding, and there is a long
line of research addressing this problem~\cite{Elliott:1997b,Claessen:99,Gill:2009,Svenningsson:2012,McDonell:2013}.
Moreover, we would like the concrete syntax of the embedded language to be as close as possible to the equivalent
expression in the host language. 

Operator and syntax overloading
help to make an embedding feel natural by providing implementations for
primitive operations such as @(+)@ or @if-then-else@ on embedded terms. Higher-order
embeddings, which make host language variables and abstraction
available to the user, are also add convenience. The higher-order embedding can then be converted to a first-order
embedding, again transparently to the user~\cite{Claessen:99,Gill:2009,McDonell:2013}.
Many deeply embedded languages do not support algebraic data types at all, or only a fixed set of built-in data
types. To understand why the challenges, in particular posed by the integration of algebraic data types, let us have a
look some concrete examples. We start with the following Haskell function:
%
\begin{haskell}
safeDiv :: Fractional a => %\textcolor{host}{a}% -> %\textcolor{host}{a}% -> %\textcolor{host}{Maybe a}%
safeDiv n d =
  if d == 0                 -- operator; conditional
     then Nothing           -- constructor
     else let r = n / d     -- operator; let-binding
           in Just r        -- constructor
\end{haskell}
%
Even though this function is quite simple, and only returns an algebraic data type as result, it
already demonstrates much of the complexity of the task.
%
An implementation of the function in an embedded language might have the type signature:
\begin{haskell}
safeDiv' :: Fractional (Exp a)
         => %\textcolor{exp}{Exp a}% -> %\textcolor{exp}{Exp a}% -> %\textcolor{exp}{Exp (Maybe a)}%
\end{haskell}
This embedded function does not take regular numeric values as input. Instead, it takes
two \emph{abstract syntax trees} of type @Exp a@, which represent a
computation that---once evaluated---will result in a numeric value of type
@a@. By reifying the program as an abstract syntax tree we have much more
freedom in how we evaluate it. The evaluation function could simply be an
interpreter in the host language, but it could also generate and compile code
for the expression.

\begin{figure}[t]
\includegraphics[width=\columnwidth]{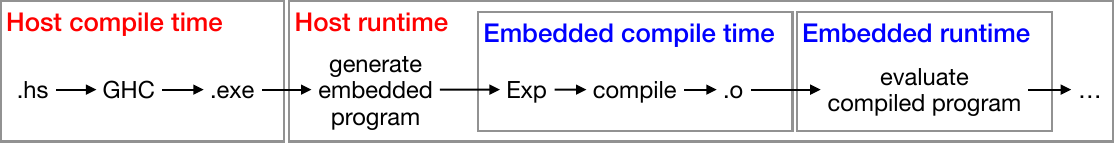}
\caption{Execution stages of a deeply embedded program}
\label{fig:phase_diagram}
\end{figure}

Fig.~\ref{fig:phase_diagram} shows an overview of the stages of
evaluation for a language deeply embedded in Haskell. The key idea is that, when
we write a program in a deeply embedded language, what we are really doing is
writing a program in the \emph{host} language which, at host program
\emph{runtime}, will generate the abstract syntax for the embedded
program (@Exp a@), compile (or interpret) that program, execute it (possibly on a
different computing device), and finally return the result of running the
embedded program back into the host program. This cycle may be repeated many
times during the execution of the host program.

This \emph{staged compilation} means that values of type @a@ are fundamentally
different from values of type @Exp a@. While we can often \emph{lift} values
from the host to the embedded language by generating appropriate abstract syntax
to represent that value, the reverse is not possible. The only way to return
expression values back to the host is by evaluating the \emph{entire} embedded program.

We use the colours of the different stages of program
execution presented in Fig.~\ref{fig:phase_diagram} to provide a visual hint of
where (or rather when) a value is available:
either during execution of the
\textcolor{host}{host} or \textcolor{exp}{embedded} program.
Effectively crossing between the different phases of embedded
program execution is the key challenge solved by this paper.

Returning to our example, the challenges to cleanly implementing 
@safeDiv'@ in a combinator-based deeply embedded language are, in order of
increasing complexity:

\paragraph{Type class overloading:}

Operations in a deeply embedded language do not directly compute values.
Instead, they combine the abstract syntax tree(s) of their argument(s) to form a
new AST that represents the result of computing that operation. In Haskell, the
division operator @(/)@, for example, is provided by the @Fractional@ type
class. If the embedded language author
provides an instance of this type class:
\begin{haskell}
instance Fractional (Exp a) where
  (/) = ...      -- construct appropriate abstract syntax
\end{haskell}
then the @(/)@ operation can be used transparently on
embedded terms, and so programming in the embedded language feels a lot like
programming in the host language. 

\paragraph{Conditionals:}

Host language conditionals cannot be used directly on embedded Boolean expressions,
since the condition needs to be of type @Bool@, not @Exp Bool@: the latter represents a
calculation whose Boolean value cannot be determined at the time of generating
the embedded program, only after that program is evaluated. Embedded languages
often work around this limitation by instead providing a \emph{function} that can be
used to express embedded conditionals:
@cond :: Exp Bool -> Exp a -> Exp a -> Exp a@. If the host language supports
overloading or rebinding of the built-in syntax to this expression language
operation, then the traditional @if-then-else@ syntax can be used in the
embedded program as well.

\paragraph{Variable bindings:}


It is important to ensure that sharing of
@let@-bound embedded expressions is preserved, and their value is not recomputed for every use of
the bound variable. 
Previous work has shown how to efficiently and
conveniently use the host language mechanisms for variable binding and
abstraction by using a higher-order embedding~\cite{Pfenning:1988}, which is
then internally converted to a first-order embedding~\cite{Atkey:2009} while
taking care to preserve~\cite{McDonell:2013} or
recover~\cite{Gill:2009,Axelsson:2012} any sharing. This conversion has to happen as the very 
first step when processing the AST to preserve sharing.  

The techniques we describe in this paper are all concerned with
the initial construction of the higher-order AST and do not interfere with sharing recovery.

\paragraph{Algebraic data types:}

Arguably, a functional language should support algebraic data types
(ADTs), both user defined as well as built-in types such as @Maybe@. This
means that we must address: (1) lifting algebraic data type values into the
embedded language; and (2) constructing, inspecting, and decomposing
embedded expressions on algebraic data types.

Lifting values of algebraic data types means the embedded language implementer
must devise a mapping of the user-extensible set of \emph{surface} types into
a fixed set of \emph{representation} types internally understood by the
embedded language. 
The  challenge with such a mapping is to do it in a way
such that implementation details of the embedding do not leak to the user, for
example via compiler error messages. Failing to do so can seriously affect the
usability of the embedding.

The difficulty is in inspecting and decomposing abstract data type values in the
embedded language, particularly for sum types. Consider the following
function:
\begin{haskell}
fromMaybe :: %\textcolor{host}{a}% -> %\textcolor{host}{Maybe a}% -> %\textcolor{host}{a}%
fromMaybe d Nothing  = d
fromMaybe _ (Just x) = x
\end{haskell}
If we want to write the equivalent of this Haskell function in the embedded language, we unfortunately cannot pattern
match directly on the second argument, for the same reason we could not use the built-in conditional operator:
we will only know which constructor was passed as argument to the
function once the expression has been evaluated; it is not available when
generating the abstract syntax of the embedded language program. Conditional
expressions can be seen as pattern matching specialised to Boolean
values, where we were able to work around this limitation by using a function
@cond@ instead. Pattern matching, however, is more general and can therefore not
be replaced by a single operator: for every pre-defined and user-defined
algebraic data type, we need the functionality to create a term in the embedded
language which expresses a @case@ distinction to be made at embedded program
execution time.


In this paper we show how we can lift the pattern matching functionality of the
host language into the embedded language, and how to integrate this feature in a
user-friendly way that is as close as possible to the syntax of pattern matching
in the host language.

As we will see, the embedded version of the @fromMaybe@ function can be
implemented almost as the non-embedded version, only the names of the constructors in the pattern need to change:
\begin{haskell}
fromMaybe' :: %\textcolor{exp}{Exp a}% -> %\textcolor{exp}{Exp (Maybe a)}% -> %\textcolor{exp}{Exp a}%
fromMaybe' d Nothing_  = d
fromMaybe' _ (Just_ x) = x
\end{haskell}
Similarily, the full embedded implementation of the @saveDiv@ function is also almost the same, apart from the name change:
\begin{haskell}
safeDiv' :: Fractional (Exp a)
         => %\textcolor{exp}{Exp a}% -> %\textcolor{exp}{Exp a}% -> %\textcolor{exp}{Exp (Maybe a)}%
safeDiv' n d =
  if d == 0                 -- operator; conditional
     then Nothing_          -- constructor
     else let r = n / d     -- operator; let-binding
           in Just_ r       -- constructor
\end{haskell}

Before we discuss what happens behind the scenes to realise this embedding, we briefly revisit pattern matching in
Haskell, as this will form the basis for our contribution.



\section{Pattern Matching in Haskell}
\label{sec:pattern_matching}


\begin{figure}[t]
\small
\centering
\begin{tabular}{lclp{5pt}l}
$var$, $x$ & ::= & \dots                        &  & Variable name         \\
$con$, $K$ & ::= & \dots                        &  & Data constructor name \\
$expr$     & ::= & \dots                        &  & Expression            \\
$pat$      & ::= & $var$                        &  & Variable pattern      \\
           & $|$ & `\textunderscore'            &  & Wildcard pattern      \\
           & $|$ & $con$ $pat_1$ \dots $pat_n$  &  & Constructor pattern   \\
           & $|$ & $expr$ `$\rightarrow$' $pat$ &  & View pattern          \\
\end{tabular}
\caption{The syntax of patterns in Haskell}
\label{fig:patterns}
\end{figure}


When describing our approach, we frequently refer to the syntax and semantics of
pattern matching in Haskell~\cite{Haskell2010} and we use its pattern synonyms
extension~\cite{Pickering:2016} to make the integration between the host and
embedded languages seamless. Therefore, we start with a brief summary of both.


Fig.~\ref{fig:patterns} gives the core syntax of patterns in Haskell.
Variable and constructor patterns are conventional, while view patterns are an extension to the basic syntax to allow
computations to be performed during pattern matching. For example:
\begin{haskell}
uncons   :: [a] -> Maybe (a, [a])
safeHead :: [a] -> Maybe a
safeHead (uncons -> Just (x, _)) = Just x
safeHead _                       = Nothing
\end{haskell}
%
%
%
By using @uncons@ inside a view pattern, matching the result of the view
function against the @Just@ constructor determines whether the first equation
for @safeHead@ succeeds or not.


\subsection{Pattern Synonyms}
\label{sec:pattern_synonyms}

Pattern synonyms~\cite{Pickering:2016} are an extension to Haskell pattern
matching that allows programmers to define new patterns.
In an \emph{implicitly bidirectional} pattern synonym 
the right-hand
side 
is used both as a pattern (when matching) and as an
expression (when constructing). This works nicely because many Haskell patterns
look like expressions, but, as we saw with view patterns, sometimes we want to
do some \emph{computation} at the same time. For example, suppose we want to
convert between rectangular and polar coordinates:
\begin{haskell}
data Cartesian = Cartesian Float Float
toPolar   :: Cartesian -> (Float, Float)
fromPolar :: Float -> Float -> Cartesian
\end{haskell}
%
Rather than define a new datatype for polar coordinates and juggling both
representations, we can convert between Cartesian and polar coordinates on
demand by writing the following \emph{explicitly bidirectional} pattern synonym:
\begin{haskell}
pattern Polar :: Float -> Float -> Cartesian
pattern Polar r a <- (toPolar -> (r, a))
  where Polar = fromPolar
\end{haskell}
There are two pieces to this declaration. First, the back-arrow
``$\leftarrow$'' indicates that @Polar@ can be used in patterns,
wherein it consists of the view function @toPolar -> (r,a)@ to first
compute the variables @r@ and @a@, which are then bound on the left. Second,
the @where@ clause specifies how @Polar@ should behave when used as an
expression, in this case by converting the given polar coordinates into
Cartesian space.


\begin{figure}[t]
\small
\centering
\begin{tabular}{lcll}
$psyn$    & ::= & `\textbf{pattern}' $lhs$ `$=$' $pat$          & Implicitly bidirectional \\
          & $|$ & `\textbf{pattern}' $lhs$ `$\leftarrow$' $pat$ & Explicitly bidirectional \\
          &     & \ \ \,`\textbf{where}' $lhs$ `$=$' $expr$     &                          \\ [0.5em]
$lhs$     & ::= & $con$ $var_1$ \dots $var_n$                   & Prefix notation          \\
$sig$     & ::= & `\textbf{pattern}' $con$ `::' $pat\_ty$       & Pattern signature        \\
$pat\_ty$ & ::= & \dots                                         & Pattern type             \\
\end{tabular}
\caption{The syntax of pattern synonym declarations}
\label{fig:pattern_synonyms}
\end{figure}





The definition in the @where@ clause is the \emph{builder}, while the
pattern after the ``$\leftarrow$'' is the \emph{matcher}. Critically,
the only required relationship between the builder and the matcher is
that their types are compatible.
One might expect that the builder and matcher have to be inverses of each other, as in
@Polar@, but this is not the case. When we define our embedded
pattern matching in Section~\ref{sec:implementation} we will make use of
this asymmetry, as both constructing \emph{and} destructing embedded terms \emph{adds}
abstract syntax to an expression. A subset of the syntax of pattern
synonyms is shown in Fig.~\ref{fig:pattern_synonyms}.




\begin{figure}[t]
\small
\noindent
\centering
\begin{tabular}{lclp{5pt}l}
$E$        & ::= & $C$                                        &  & Constant        \\
           & $|$ & $x$                                        &  & Variable        \\
           & $|$ & \textbf{let} $var$ $=$ $E$ \textbf{in} $E$ &  & Let binding     \\
           & $|$ & $\lambda x.\ E$                            &  & Abstraction     \\
           & $|$ & $E$ $E$                                    &  & Application     \\
           & $|$ & $T$                                        &  & Tuples          \\
           & $|$ & $\pi$ $E$                                  &  & Projection      \\
           & $|$ & $trace$ $E$                                &  & Pattern match   \\
           & $|$ & $E$ [ $E_1$ \dots $E_n$ ]                  &  & Case expression \\
$T$        & ::= & $()$                                       &  & Unit            \\
           & $|$ & $E$                                        &  & Expression      \\
           & $|$ & $T$ $T$                                    &  & Pair            \\
$\pi$      & ::= & \dots                                      &  & Projections     \\
$C$        & ::= & \dots                                      &  & Constants       \\
$var$, $x$ & ::= & \dots                                      &  & Variable name   \\
$trace$    & ::= & \dots                                      &  & Match trace (\S\ref{sec:nested_patterns}) \\
\end{tabular}
\caption{The grammar of our embedded language}
\label{fig:language}
\end{figure}

\subsection{Semantics of Pattern Matching}
\label{sec:pattern_semantics}


In this section we summarise the semantics of pattern matching in
Haskell~\cite{Haskell2010,Pickering:2016}. When matching against a value, there
are three possible outcomes: the match \emph{succeeds} (binding some variables);
the match \emph{fails}; or the match \emph{diverges} (does not terminate or
terminates with an error). Failure is not necessarily an error:
if the match fails in a function defined by multiple equations, or a case
expression with multiple alternatives, the next equation (or alternative) is tried.

Matching a pattern $p$ against a value $v$ is given by cases which
depend on the form of the pattern. If it is
\begin{enumerate}[leftmargin=*]
  \item a variable $x$ the match succeeds, binding $x$ to
    the value $v$.

  \item a wildcard `\textunderscore' the match succeeds, binding
    nothing.

  \item a view pattern @f -> p@ then evaluate @f v@. If
    evaluation diverges, the match diverges, otherwise match the resulting value
    against $p$.

  \item of the form $(P\ p_1 \ldots p_n)$, where $P$
    is a data constructor, or a pattern synonym defined by
    $P\ x_1 \ldots x_n = p$ or $P\ x_1 \ldots x_n \leftarrow p$, then 
    \begin{enumerate}
      \item if the value $v$ diverges, the match diverges.
      \item if the value is of the form $(P'\ v_1 \ldots v_m)$ where $P \neq P'$
        the match fails.
      \item \label{enum:sub_patterns} if the value is of the form $(P\ v_1
        \ldots v_n)$ then we match the sub-patterns from left-to-right ($p_1$
        against $v_1$ and so on). If any of the patterns fail (or diverge) then
        the whole pattern match fails (or diverges). The match succeeds if all
        sub-patterns succeed.
    \end{enumerate}

  %
  %
  %
  %
  %
\end{enumerate}

The main point to note is that for a pattern match to succeed,
matching on the constructor or pattern synonym---\emph{as well as all of its
sub-patterns}---must succeed.
The need to handle nested patterns complicates our task considerably.
We will return to this point in Section~\ref{sec:nested_patterns}.


\subsection{Cosmetic Shortcomings of Pattern Synonyms}
\label{sec:cosmetic_shortcomings}

Using pattern synonyms for working with embedded data
types provides a significant usability improvement for the embedded language
user (\S\ref{sec:evaluation}). However,
there are two cosmetic shortcomings of pattern synonyms in Haskell:
\begin{itemize}[leftmargin=*]
  \item Pattern synonyms exist in the same name space as data constructors, so
    we cannot use the same name for a constructor and its corresponding
    embedded pattern synonym. Our convention is to form the
    pattern synonym name by adding a trailing underscore to the constructor
    name.

  \item It is not possible to overload built-in syntax, in particular for
    tuples. Instead we use the pattern synonym name @T2@ for pairs @(,)@, @T3@
    for triples @(,,)@, and so on.
\end{itemize}

\section{Generic Expression Language}
\label{sec:expression}

We explain our technique for embedding algebraic data types using the 
language shown in Fig.~\ref{fig:language}. It offers only a
few operations: lifting primitive values into the language, construction and
access to pairs, and case expressions.
We elide discussion of bindings and application, as it is standard and does not
affect the technique.
In this section we present a representation for this core language, and show how
user-defined algebraic data types can be minimally encoded in it.

\begin{figure}[t]
\vspace{-\medskipamount}
\begin{lstlisting}
data Exp a where
  Const  :: EltR a -> Exp a
  Tuple  :: Tuple (TupleR t) -> Exp t
  Prj    :: TupleIdx (TupleR t) e -> Exp t -> Exp e
  Roll   :: Exp a -> Exp (Rec a)
  Unroll :: Exp (Rec a) -> Exp a
  Match  :: TraceR (EltR a) -> Exp a -> Exp a
  Case   ::
    Exp a -> [(TraceR (EltR a), Exp b)] -> Exp b

data Tuple t where
  Unit   :: %\phantom{$\leftarrow$}%%\phantom{$\leftarrow$}%                  Tuple ()
  Exp    :: Exp a            %\phantom{$\leftarrow$}%-> Tuple a
  Pair   :: Tuple a -> Tuple b -> Tuple (a, b)

data TupleIdx s t where
  PrjZ   ::               %\phantom{$\leftarrow$}%TupleIdx t      t
  PrjL   :: TupleIdx l t -> TupleIdx (l, r) t
  PrjR   :: TupleIdx r t -> TupleIdx (l, r) t

newtype Rec a = Rec a
\end{lstlisting}
\caption{The core AST of our generic expression language}
\label{fig:AST}
\end{figure}

The terms of this language can be implemented in Haskell via the data type shown
in Fig.~\ref{fig:AST}, which is defined as a generalised algebraic data type
(GADT)~\cite{Jones:2004,Vytiniotis:2016} that adds a \emph{type level} index to
the expression syntax tree. Defining @Exp t@, rather than @Exp@, denotes that
evaluating the expression yields a value of type @t@, which is checked during
compilation of the host program. This intrinsically typed representation ensures
that embedded language programs written by the user, as well as AST
transformations written by the language implementer, are type correct by
construction. While choosing a typed representation is a largely orthogonal
property of the program representation and makes the techniques we present here
more involved, it is well worth the cost for complex embedded
languages~\cite{Chakravarty:2011,McDonell:2013,McDonell:2015}.

The user-facing aspect of the embedded language are terms of type @Exp a@. These
terms are parameterised by the \emph{surface} type @a@ of the value that term
represents---that is, the user's view of a data type---just as in the previously
discussed examples.
The type class~\cite{Peterson:1993kc,Hall:1996ez} @Elt@ characterises the
\emph{extensible} set of surface types expressible in the language. Its
associated type~\cite{Chakravarty:2005dx,Schrijvers:2008ir}
@EltR@ maps the surface type to a corresponding internal representation type in terms
of the \emph{closed} set of primitive types of the language, unit
@()@, pair @(,)@, and @Rec@, with @fromElt@ and @toElt@ converting between the
two.
\begin{haskell}
class Elt a where
  type EltR a
  fromElt :: a -> EltR a
  toElt   :: EltR a -> a
  traceR  :: [TraceR (EltR a)]  -- Explained in \S\ref{subsubsec:embeddingPM}
\end{haskell}
%
%

%
%

%
It is up to the designer of the embedded language to decide the set of primitive
types supported by their language, but this choice
does not affect the techniques presented here. An embedded language with
built-in support for product types and mutable references, for example, is
equally amenable to our technique, which is to enable \emph{user-defined}
algebraic data types built from nested pairs of the types in this set.


Instances of the @Elt@ class are automatically derivable via GHC 
Generics~\cite{Gibbons:2006} (for non-recursive data types).
We will show some concrete examples of how algebraic data types are encoded in
this representation in the following sections.

\subsection{Representation of Product Types}
\label{sec:product_types}

Data types in our internal representation are either constant values introduced
by the constructor @Const@, which lifts supported values from the host language
into the embedded language, or nested tuples thereof. Nullary and binary tuples
are represented by the constructors @Unit@ and @Pair@ respectively, with @PrjL@
and @PrjR@ to project the first and second component of a pair.

We represent surface level product types by isomorphic nested pair types. As
mentioned previously this mapping from surface to representation type is
accomplished via the type class @Elt@ and its associated type @EltR@. For
example, the surface triple type @(a, b, c)@ is represented internally by the
type @((((), EltR a), EltR b), EltR c)@.
Primitive types have identity representations.
%
%
Similarly, user-defined product types such as:
\begin{haskell}
data V2 a  = V2 a a
data Point = Point Float Float
\end{haskell}
are mapped to the following nested pair types:
\begin{haskell}
EltR (V2 a) = (((), EltR a), EltR a)
EltR Point  = (((), Float), Float)
\end{haskell}

In addition to the associated type @EltR@ which \emph{deeply} maps a surface
type into nested pairs of primitive types, the type class @IsTuple@ and
associated type @TupleR@ perform this mapping to binary pairs at only a
\emph{single} level. For example, the surface triple type @(a, b, c)@ is
represented by the nested pair type @((((), a), b), c)@.
%
%
As with the type class @Elt@, instances of the class @IsTuple@ and its
associated type @TupleR@ are derived automatically. 
These two classes define how to map an
arbitrarily nested data type into its constituent primitive values
(@Elt@),\footnote{Deeply converting a surface type to nested pairs of primitive
types is necessary for external code generation, for example.}
and how to project components out at each level of that nesting (@IsTuple@).
%
Continuing the example: 
%
\begin{haskell}
TupleR (V2 a) = (((), a), a)
TupleR Point  = (((), Float), Float)
\end{haskell}
A value of type @Point@ can be lifted from the host language into
the embedded language by representing it with the following abstract
syntax term:\footnote{\label{foot:liftPoint}We could also write
\lstinline[style=footnote]{liftPoint = Const . fromElt} but the chosen exposition allows us to
demonstrate how abstract syntax fragments are combined to produce terms of product
type.}
\begin{haskell}
liftPoint :: %\textcolor{host}{Point}% -> %\textcolor{exp}{Exp Point}%
liftPoint (Point x y) = Tuple $
  Unit `Pair` Exp (Const x) `Pair` Exp (Const y)
\end{haskell}
while extracting the $x$-component of an embedded @Point@ is represented by:
\begin{haskell}
xcoord :: %\textcolor{exp}{Exp Point}% -> %\textcolor{exp}{Exp Float}%
xcoord p = Prj (PrjL (PrjR PrjZ)) p
\end{haskell}

As we can see in @xcoord@, the decomposition of embedded values \emph{adds} embedded
language terms. To extract the individual fields of a constructor, we cannot pattern match
directly on the @Pair@ term.
\footnote{For example, the function \lstinline[style=footnote]{liftPoint} as defined in the body
text or in footnote~\ref{foot:liftPoint} produce different abstract syntax trees
to represent the same value, and we would like that our
\lstinline[style=footnote]{xccord}
function work given any type correct term.}
We return to this example in Section~\ref{sec:products}.

\subsection{Representation of Sum Types}
\label{sec:sum_types}

In contrast to product types, which can be mapped directly to nested
pairs, it is not obvious what the internal representation of sum types should
be. 
%
%
%
%
%
To implement embedded function @fromMaybe'@ from Section~\ref{sec:embedded_languages}
we require: (a) a
representation for values of type @Exp (Maybe a)@; and (b) a way to distinguish
whether the embedded term represents the @Nothing@ or @Just@ constructor.
Meeting the second requirement
in such a manner that
\emph{embedded} terms can be used in pattern matching in the \emph{host}
language is the crux of the challenge.
%

Recall that
during the runtime of the host language program the
AST representing the embedded program 
will
be generated, compiled, and executed, 
before returning the
result back into the host program. 
%
Thus we can only know which pattern match will succeed \emph{after}
the abstract syntax for the program has been generated and the
argument 
evaluated. But we can only generate this abstract syntax 
by
exploring the right-hand-side of the function, which requires matching on the
argument and must surely happen \emph{before} executing the program. Pattern
matching on embedded sum types therefore would require mixing two distinct phases of
program execution: host program runtime (which performs the pattern matching)
and embedded program runtime (when the values to be matched are available).

We break this cycle in two parts. First, by having a \emph{uniform}
representation for the top-level constructor of each alternative in an embedded
sum type; and second, by the method we will use to inspect terms of this type that we
will introduce in Section~\ref{sec:sums}.
For example, the following familiar data types are mapped to this uniform representation as:
%
\begin{haskell}
EltR Bool         = (%\hostTypeTAG%, ())
EltR (Maybe a)    = (%\hostTypeTAG%, ((), EltR a))
EltR (Either a b) = (%\hostTypeTAG%, (((), EltR a), EltR b))
\end{haskell}

The left component of the pair is a tag---such as @Int@---which indicates which
constructor of the sum type the value represents,
while the right component of the pair
contains the values associated with every constructor of that type.%
\footnote{Alternative representations, such as a tagged
pointer to an object containing only the data associated with that particular
constructor are of course possible, but that choice is orthogonal to the
contribution of this work. We chose a flattened representation of
primitive types to (a) simplify the internal language, eliminating the need to
deal with references; and (b) because our motivating use case is for
data-parallel arrays, for which unboxed data without pointers is the best
representation for performance applications and constrained devices such as
GPUs~(\S\ref{sec:evaluation}).}
%

We can lift a value of type @Maybe Float@ from the host 
into the embedded language with the following abstract syntax:
%
\begin{haskell}
liftMaybe :: %\textcolor{host}{Maybe Float}% -> %\textcolor{exp}{Exp (Maybe Float)}%
liftMaybe Nothing  = Tuple $
  Exp (%\hostTag{0}%) `Pair` (Unit `Pair` Exp undef)
liftMaybe (Just a) = Tuple $
  Exp (%\hostTag{1}%) `Pair` (Unit `Pair` Exp (Const a))
\end{haskell}
where the function @undef@ generates the abstract syntax for an undefined value
of the required type.%
\footnote{Values associated with the non-represented alternative contain
undefined values, but this is safe as our automatically generated pattern
synonyms will never inspect those values (\S\ref{sec:sums}).
}
Note that
this is not the Haskell term @undefined@, rather it is an embedded term that
represents a real value with an unspecified bit pattern.%
\footnote{Since the type class \lstinline[style=footnote]{Elt} tells us how to map every
supported surface type into nested pairs of primitive types, we can always
generate an unspecified value of the appropriate type, for example by choosing
zero everywhere.}

Even with this uniform representation we cannot
implement pattern matching by directly inspecting the tag value of a term.
Although @liftMaybe@ produced tags as constant values, in general
these are ASTs representing arbitrarily
complex expressions. We address this problem in Section~\ref{sec:implementation}.

\subsection{Representation of Recursive Types}
\label{sec:recursive_types}


Finally, we outline how to encode terms representing recursive data types.
%
%
We take the standard iso-recursive approach of treating the recursive type and
its one-step unfolding as different, but isomorphic~\cite{TAPL}. For example,
for the @List@ data type we obtain the following representation type, where
@Rec@ indicates the location of the recursive type:
%
\begin{haskell}
data List a = Nil | Cons a (List a)
EltR (Rec a)  = Rec a
EltR (List a) = (TAG, (((), EltR a), Rec (List a)))
\end{haskell}
%
In expressions of recursive type, the recursive term 
 is wrapped in the @Roll@ instruction (e.g., the @Cons@ constructor at
the term representing the list tail), and @Unroll@ is used to inspect a
recursive term.
Although this may seem burdensome, in practice these annotations are hidden by
our embedded pattern synonyms, which we describe in the following section.

\section{Implementation}
\label{sec:implementation}

Now that we have introduced our minimal embedded language and demonstrated
how to encode algebraic data types in it, we show how we can use pattern synonyms to conveniently
work with embedded terms of algebraic data type, and how to integrate this with
pattern matching in the host language.

\subsection{Embedded Products}
\label{sec:products}


Recall from Section~\ref{sec:product_types} the data type @Point@ and its lifting into our
expression language. In the host language we pattern match on the constructor
@Point@ to access the $x$- and $y$-components.
%
To achieve this same functionality in the embedded language, we need a
pattern synonym with type:%
\footnote{While we make use of the \emph{type} \lstinline[style=footnote]{Point}, we do not use
the host language \emph{constructor} \lstinline[style=footnote]{Point}. This is precisely what
the pattern synonym provides a replacement for in embedded language code.}
\begin{haskell}
pattern Point_ :: %\textcolor{exp}{Exp Float}% -> %\textcolor{exp}{Exp Float}% 
               -> %\textcolor{exp}{Exp Point}%
\end{haskell}

As both constructing a term of type @Exp Point@ as well as destructing these
terms to access the stored components amounts to adding new abstract
syntax, we require an explicitly bidirectional pattern synonym.
The builder function, which specifies how the
synonym should behave as an expression, is straightforward:
\begin{haskell}
buildPoint :: %\textcolor{exp}{Exp Float}% -> %\textcolor{exp}{Exp Float}% -> %\textcolor{exp}{Exp Point}%
buildPoint x y = Tuple $
  Unit `Pair` Exp x `Pair` Exp y
\end{haskell}
as is the matcher function, which extracts each component:
\begin{haskell}
matchPoint :: %\textcolor{exp}{Exp Point}% -> %\textcolor{host}{(}%%\textcolor{exp}{Exp Float}%%\textcolor{host}{,}% %\textcolor{exp}{Exp Float}%%\textcolor{host}{)}%
matchPoint p = ( Prj (PrjL (PrjR PrjZ)) p
               , Prj (PrjR PrjZ) p )
\end{haskell}
Note how the matcher returns its two expression language fragments in a regular Haskell pair.
The completed pattern synonym is then:
\begin{haskell}
pattern Point_ x y <- (matchPoint -> (x, y))
  where Point_ = buildPoint
\end{haskell}
We can now use pattern matching in the embedded language in much the same way as in the host language, for example:
\begin{haskell}
addPoint :: %\textcolor{exp}{Exp Point}% -> %\textcolor{exp}{Exp Point}% -> %\textcolor{exp}{Exp Point}%
addPoint (Point_ x%$_1$% y%$_1$%) (Point_ x%$_2$% y%$_2$%) =
  Point_ (x%$_1$% + x%$_2$%) (y%$_1$% + y%$_2$%)
\end{haskell}

Furthermore, due to our uniform representation, defining embedded pattern synonyms for all
product types works in the same
way. We can abstract this procedure into the following polymorphic pattern synonym:
\begin{haskell}
pattern Pattern :: IsPattern s r => r -> Exp s
pattern Pattern r <- (matcher -> r)
  where Pattern = builder

class IsPattern s r where
  builder :: r -> Exp s
  matcher :: Exp s -> r
\end{haskell}

The class @IsPattern@ states that we can consider values of type @s@ (the user's
data type) as embedded language terms by representing them as we would terms of
type @r@. We can write instances for this class that will cover any type @r@ whose
\emph{representation} type @EltR r@ is equivalent to (@~@) the representation
type of @s@. For example, the following instance covers \emph{any} type whose
representation is isomorphic to a pair:
\begin{haskell}
instance ( EltR s ~ EltR (a, b)
         , TupleR s ~ TupleR (a, b) )
    => IsPattern s (Exp a, Exp b) where
  builder (x, y) = Tuple $
    Unit `Pair` Exp x `Pair` Exp y
  matcher s = ( Prj (PrjL (PrjR PrjZ)) s
              , Prj (PrjR PrjZ) s )
\end{haskell}
Note the similarity of this instance declaration
to our specialised functions
@buildPoint@ and @matchPoint@.
The embedded language author defines appropriate instances of this class
\emph{once}---for pairs @(,)@, triples @(,,)@, and so forth---and the embedded language
user can then define pattern synonyms for their (product) data types without needing to
explicitly write the required builder and matcher functions.
Our example can now be defined simply as:
\begin{haskell}
pattern Point_ x y = Pattern (x, y)
\end{haskell}
%

\subsection{Embedded Sums}
\label{sec:sums}


We demonstrated how pattern synonyms can be used to
provide pattern matching for product types in the embedded language. This was
possible because pattern synonyms for product types never need to inspect their
argument. In this section we discuss the treatment of \emph{sum} data
types---types with more than one constructor.

In Section~\ref{sec:sum_types} we presented the representation for sum
data types in the embedded language, and the function @liftMaybe@ was an
example of how to construct expressions in this form. The challenge is how to
treat these values when they are used in patterns. 
Consider the following embedded function:
\begin{haskell}
simple :: %\textcolor{exp}{Exp (Maybe Int)}% -> %\textcolor{exp}{Exp Int}%
simple Nothing_  = 0
simple (Just_ x) = x
\end{haskell}
This gets de-sugared by the host language compiler to a (host language) lambda abstraction of the form

\begin{haskell}
simple = \p -> ...
\end{haskell}

The lambda abstraction needs to construct a suitable AST. For that, it
cannot inspect its argument @p@ as it does not have the actual value that this
function will be applied to---that will only be available much later, during
embedded program execution. It can only embed @p@ as it is into the AST.

To generate the abstract syntax for this function, we need a
matcher function which is essentially able
to traverse below the lambda abstraction for each equation in the case
distinction and extract the corresponding continuation; either the constant value zero, or an expression
which extracts the value from the @Just@ constructor.

The key idea for the matcher function is that we need to evaluate the function
@simple@ \emph{twice} to construct
the corresponding embedded program. While we don't have the actual value that this
function will be applied to, we can craft a dummy argument that forces a
specific pattern in the \emph{host} language to succeed. This way, we will be able to
explore each of the right-hand-sides in turn, and construct a program with an
\emph{embedded} case statement from those fragments.

For this to work, the construction of the AST for the
embedded program has to be a pure function. This is straightforward in our example host
language, Haskell, as the type system ensures there are no side effects that
might make this technique unsafe, but in less idealistic languages this is the
responsibility of the embedded language developer. The restriction of
the AST construction to pure functions  does not preclude the
embedded language itself from containing side effects, however.

\subsubsection{First Steps}

Let us illustrate the problem by first writing the necessary
matcher function manually, before we discuss in
Section~\ref{subsubsec:embeddingPM} this how can be generated automatically.
The matcher function has the following type:
\begin{haskell}
match_simple%$_1$% :: %\textcolor{exp}{Exp (Maybe Int)}% -> %\textcolor{exp}{Exp Int}%
\end{haskell}
This function needs to perform the procedure we outlined, evaluating the function
@simple@ twice to extract the @Nothing@ and @Just@ continuations. It
has to return a new embedded function of the same type as @simple@, but where
the case statement has been lifted from the \emph{host} language into the
\emph{embedded} language. This is where we need our two new
language terms: @Match@ and @Case@ (Fig.~\ref{fig:AST}).
%

Intuitively, @Match@ bundles a host language pattern term (its first
argument) for each alternative with the actual function (@simple@ in
our example). The first argument of @Case@ is the embedded language
term we match on, and a list of expressions constructed via @Match@,
paired with exactly the same trace stored in the @Match@.  These constructors
help us perform the pattern matching and construct the embedded case expression,
respectively. We can implement the matching function for our example as:
\begin{haskell}
match_simple%$_1$% = \p ->
  let rhs_nothing = simple (Match %\embeddedTag{0}% p)
      rhs_just    = simple (Match %\embeddedTag{1}% p)
  in
  Case p [ (%\embeddedTag{0}%, rhs_nothing)
         , (%\embeddedTag{1}%, rhs_just) ]
\end{haskell}
The constructor @Match@ will be consumed by the embedded pattern synonym
(it will not be present in the abstract syntax of the embedded program)
and is parameterised by a \emph{trace} that will cause a particular host language
pattern match to succeed. We give the precise definition of \embeddedTag{}
in Section~\ref{sec:nested_patterns}.
This produces the following abstract syntax for the function @simple@:
\begin{haskell}
\p -> Case p
        [ (%\embeddedTag{0}%, Const 0)
        , (%\embeddedTag{1}%, Prj (PrjR (PrjR PrjZ)) p) ]
\end{haskell}
%
The interesting part of the @match_simple@ implementation is the use
of the @Match@ constructor, which is the ``dummy argument'' we use to
explore the pattern match alternatives.  To see how this works, we turn our
attention to the definition of the embedded pattern synonyms for @Maybe@.

\subsubsection{Embedding \texttt{Maybe}}

As an example, consider defining the embedded pattern synonym for the @Just@
constructor.
The builder function is straightforward; it bundles the supplied term together
with a concrete tag value indicating which constructor the argument is
associated with:
%
\begin{haskell}
buildJust :: %\textcolor{exp}{Exp a}% -> %\textcolor{exp}{Exp (Maybe a)}%
buildJust x = Tuple $
  Exp (%\hostTag{1}%) `Pair` (Unit `Pair` Exp x)
\end{haskell}
The corresponding matcher function is more interesting. Let us first look
at its type signature:
\begin{haskell}
matchJust :: %\textcolor{exp}{Exp (Maybe a)}% -> %\textcolor{host}{Maybe (}%%\textcolor{exp}{Exp a}%%\textcolor{host}{)}%
\end{haskell}
If the return value is @Nothing@ this indicates that the pattern match
failed, and if it is
@Just@ that the pattern match succeeded. Moreover, this type states that, given a
term in the \emph{expression} language of type @Maybe a@, it may return in the
\emph{host} language an expression of type @Exp a@. Notice how the result of this
function is required at host runtime, before the value of the embedded
term will be known. 

Instead of pattern matching against the value of the argument (which we do not
yet have) or against the argument AST itself (which may be arbitrarily complex)
we match against a known value: our term wrapped in the @Match@ constructor.
It is safe to inspect the argument expression only in this case
because our @Match@ constructor is ephemeral. That is, it is not part of the program the
user writes, rather the embedded language author inserts and consumes it during
embedded program construction whenever embedded pattern matching is performed,
as sketched in @match_simple@$_1$.%
\footnote{Since \lstinline[style=footnote]{Match} is used only when constructing the AST of the
embedded program, the trace information it contains could be encoded as an
annotation over the AST rather than as a term in it. For example, in Accelerate
the term language is defined using open recursion, so we could include this
extra information in the fixed point. This is a technique we use in other parts
of the compiler. We did not use that method here to (a) simplify the
explaination; and (b) because that makes a \emph{global} change to the data
type(s) the compiler deals with, rather than introducing only \emph{local}
modifications to how terms are handled in a few key areas. Moreover, this
constructor is not needed after we translate from the higher-order to
first-order representation in the first phase of our compiler, so this simpler
approach is also more appropriate in our particular use case.}
%
Here is a possible (not yet complete) implementation:
%
\begin{haskell}
matchJust%$_1$% x = case x of
  Match %\embeddedTag{1}% a -> Just
                        (Prj (PrjR (PrjR PrjZ)) a)
  Match _        %\;%_ -> Nothing
\end{haskell}
%
If the argument is the @Match@ term with the constructor trace corresponding to
the alternative we are interested in (here \embeddedTag{1}), then we
signal to the host language pattern matcher that the match was a success by
returning the abstract syntax that can be used to access the value
stored in the embedded @Just@ in the success continuation. If the @Match@ term
does not have the constructor trace we are interested in,
we signal to the host language that pattern matching failed by returning
@Nothing@. If the argument is not a @Match@ term, this corresponds to a usage
error, in which case we return an informative error message.
In the following section we discuss what this trace structure should be in order
to support the required behaviour, and complete the definition of @matchJust@.



\subsection{Nested Patterns and $\langle$\texttt{TRACE}$\rangle$}
\label{sec:nested_patterns}

Now that we have looked at the key idea behind our approach,
let us see how this would work for nested pattern matching.
Consider the following example:
\begin{haskell}
nested%$_1$% :: %\textcolor{exp}{Exp (Maybe Bool)}% -> %\textcolor{exp}{Exp Int}%
nested%$_1$% Nothing_       = 0
nested%$_1$% (Just_ False_) = 1
nested%$_1$% (Just_ True_)  = 2
\end{haskell}
The semantics of pattern matching (\S\ref{sec:pattern_semantics}) dictates
that for a pattern match to succeed, matching on the pattern
$P$ as well as \emph{all} of the sub-patterns of the
pattern must succeed. We do not receive
information from the host language pattern matcher as to \emph{why} a pattern
succeeded (did it match on the pattern we are interested
in, or the wildcard pattern `\textunderscore'?) and conversely if it failed, we
do not have the opportunity to recover from that failure (for example, by
only at that point
attempting sub-pattern matches).

To support nested pattern matching, we therefore need a representation of the
\embeddedTag{} which specifies a complete pathway of pattern matching through
all of the sub-patterns supported by this type. We use the following data type as both a
witness to the structure of a term, and to denote the alternative chosen at each
sub-pattern match:
\begin{haskell}
data TraceR a where
  TraceRunit :: TraceR ()
  TraceRprim :: PrimType a -> TraceR a
  TraceRpair :: TraceR a -> TraceR b 
             -> TraceR (a, b)
  TraceRrec  :: TraceR (EltR a) -> TraceR (Rec a)
  TraceRtag  :: TAG -> TraceR a -> TraceR (TAG, a)
\end{haskell}
%
The first four constructors of this data type act as witnesses to the structure
of the representation type of the term.
The last constructor
indicates the position of a concrete tag value used to discriminate
alternatives of a sum data type.
For example, the representation type of the surface type @Maybe Bool@ is:
\begin{haskell}
EltR (Maybe Bool) = (TAG, ((), (TAG, ())))
\end{haskell}
And the value @Just False@ has the following trace:
\begin{haskell}
TraceRtag 1 `TraceRpair` (TagRunit
  `TraceRpair` (TraceRtag 0 
    `TraceRpair` TagRunit))
\end{haskell}

The trace structure allows us to propagate an appropriate sub-pattern trace to
every field of the constructor, enabling the host language
to explore nested pattern matches. We can now complete the matcher function for
the @Just@ constructor:
%
\begin{lstlisting}[style=haskell,linewidth=2\textwidth]
matchJust :: %\textcolor{exp}{Exp (Maybe a)}% -> %\textcolor{host}{Maybe (}%%\textcolor{exp}{Exp a}%%\textcolor{host}{)}%
matchJust x = case x of
  Match (TraceRtag 1 (TraceRpair TraceRunit s)) a
    -> Just (Match s (Prj (PrjR (PrjR PrjZ)) a))
  Match _ _ -> Nothing
\end{lstlisting}
Adding the @Match@ term in the success continuation ensures that nested patterns
are in the right form for the embedded pattern synonyms whenever the host
language pattern matcher attempts to match on the constructor sub-patterns. The
complete embedded pattern synonym is:
\begin{haskell}
pattern Just_ :: %\textcolor{exp}{Exp a}% -> %\textcolor{exp}{Exp (Maybe a)}%
pattern Just_ x <- (matchJust -> Just x)
  where Just_ = buildJust
\end{haskell}
Finally, notice the pattern match on the result of the view function
@matchJust@: this is how the pattern synonym signals to the host pattern matcher
the success or failure of the embedded pattern match.

Although this encoding may seem complex, it is necessary for supporting nested
pattern matching while also ensuring type safety of the embedded language.
Moreover (a) the trace structure can be removed from the final embedded language
encoding to use nested case statements directly on literal @TAG@ values
(\S\ref{sec:nested_case}); and (b) these pattern synonyms are generated
automatically using TemplateHaskell~\cite{Sheard:2002}. The embedded language
user simply writes the following to have their data type available for use in
embedded code:
\begin{haskell}
mkPattern ''Maybe
\end{haskell}
%



\subsection{Embedded Recursive Types}
\label{sec:recursive}

Now that we have introduced embedded pattern synonyms for nested sum and product
types, we discuss treatment of recursive types. This works similarly to what we
have seen so far, with the addition that we must also insert a @Roll@ or
@Unroll@ term to decorate the recursion point. However, there is one key
difference. Consider the following function:
\begin{haskell}
nested%$_2$% :: %\textcolor{exp}{Exp (List Int)}% -> %\textcolor{exp}{Exp Int}%
nested%$_2$% (Cons_ _ (Cons_ _ Nil_)) = 1
nested%$_2$% _                        = 0
\end{haskell}
In order to extract the right-hand-side we evaluate the function multiple times,
providing the @Match@ constructor as argument with a different \embeddedTag{}
each time. However, the function @nested@$_2$ might have any number of
specialisations for lists of different lengths, and we receive no feedback as to
why a pattern match succeeds or fails, so how do we decide when to stop
searching for more unique right-hand-sides?

Although continuing to some arbitrarily chosen depth will catch many common
cases---for example, matching on a single-element list---there is in general no
solution to this limitation of pattern synonyms in Haskell. To avoid this
problem, our embedded pattern synonyms do not provide sub-pattern matches
against recursive types---thereby leading to a runtime error if the user
attempts to do so---and instead requires them to inspect any recursive terms
within a separate @case@ statement.

\subsection{Embedding Pattern Matching}
\label{sec:match}
\label{subsubsec:embeddingPM}

With our newly defined embedded pattern synonyms, we now only need some way to
use them such that the user can---as seamlessly as possible---reuse the
host-language pattern matching facilities inside their embedded
program. That is, the user should not have to write the matching functions
manually, as we did in @match_simple@$_1$. Instead, we provide a
higher-order function @match@ which does that automatically, such
that our example @match_simple@$_1$ becomes simply:
\begin{haskell}
match_simple%$_2$% :: %\textcolor{exp}{Exp (Maybe Int)}% -> %\textcolor{exp}{Exp Int}%
match_simple%$_2$% = match simple
\end{haskell}

We can generalise the procedure we followed to write the code for @match_simple@$_1$ into the following steps:
\begin{enumerate}[leftmargin=*]
  \item \emph{Enumerate traces:} For an expression of type @Exp a@, which
    will be the scrutinee of a @case@ statement or pattern match, we enumerate
    all possible pattern matches at the type @a@. For example, for the type
    @Maybe Bool@ there are three possibilities: @Nothing@, @Just False@, and
    @Just True@. These trace alternatives are represented by the @TraceR@ structure
    and generated by the @traceR@ function from the @Elt@ typeclass.

  \item \emph{Collect pattern alternatives:} Given a function @f@ of type
    @Exp a -> b@, we apply @f@ to each trace alternative generated from the
    previous step to extract the success continuation associated with
    that tag. Apply steps 1 and 2 recursively on @b@ for $n$-ary functions.

  \item \emph{Case introduction:} The right hand sides extracted by the previous
    step are combined into an embedded @Case@ term. At this point we can
    reintroduce nested case statements and default cases to remove redundant
    branches.
\end{enumerate}


\newcommand{\cref}[1]{\CircledText{\small\ref{#1}}}

We will use the @match_simple@$_2$ example to explain how the implementation shown
in Fig.~\ref{fig:matching} generates embedded pattern matches in two phases.
We have that:
%
\begin{haskell}
match simple = mkFun (mkMatch simple) id
\end{haskell}

First, @mkFun@ collects all of the arguments to the function as a typed
list of expressions of type @Args@.
This argument list is collected by building the continuation @k@, consuming the
arguments to the function left-to-right, one binder at a time~\cref{c:mkFunR}.
Once the function is fully saturated this structure is then passed to the
function @f@ to construct the final embedded term~\cref{c:mkFun0}. In
our example, at this point we have the argument list @x :-> Result@, where @x@
was the single argument of type @Exp (Maybe Int)@ passed to the function, and
@f@ is the function @mkMatch simple@.

Second, @mkMatch@ lifts a function over expressions (such as @simple@) to a 
corresponding function that expects its arguments stored in this @Args@ list.
For each argument in this list it generates all of
the pattern alternatives and combines them into an embedded @Case@ term,
applying this to the function @f@ as it goes to build the final expression. In
our example, at~\cref{c:mkMatchR} we have that @f@ is the function @simple@ and @x@
the single argument to that function. 

It is important to first check whether the argument @x@ is already in @Match@
form~\cref{c:nested}, which can arise if @match@ is applied multiple times.
This check prevents generating an exponential number of @Case@ terms in the
embedded code. For every @trace@
supported by this type~\cref{c:traceR}---in this example corresponding to
@Nothing@ and @Just@---we wrap the argument @x@ in the proxy @Match@
term~\cref{c:match}, enabling the corresponding embedded pattern match to
succeed. Recursing on the remaining arguments extracts all of the equations of
the argument function @f@~\cref{c:rhs}. Finally, we introduce the embedded
@Case@ term on the scrutinee @x@ and list of success continuations
@rhs@~\cref{c:case}. If the type @e@ has a single trace alternative---primitive types such as
@Int@ and data types with a single constructor---this can be
elided~\cref{c:scalar}. Note that the list @rhs@ is in the order defined by the
@traceR@ function, \emph{not} the order that these equations appear in the user's
source program.

\begin{figure}[t]
\begin{haskell}
match :: Matching f => f -> f
match f = mkFun (mkMatch f) id

data Args f where
  (:->)  :: Exp a -> Args b -> Args (Exp a -> b)
  Result :: Args (Exp a)

class Matching a where
  type Result a
  mkFun   :: (Args f -> Exp (Result a))
          -> (Args a -> Args f) -> a
  mkMatch :: a -> Args a -> Exp (Result a)

instance Elt r => Matching (Exp r) where
  type Result (Exp r) = r
  mkFun f k        = f (k Result) %\hspace*{\fill}%-- {\slshape\cstep\label{c:mkFun0}}
  mkMatch e Result = e

instance (Elt e, Matching r)
    => Matching (Exp e -> r) where
  type Result (Exp e -> r) = Result r
  mkFun f k = \x ->
    mkFun f (\xs -> k (x :-> xs)) %\hspace*{\fill}%-- {\slshape\cstep\label{c:mkFunR}}
  mkMatch f (x :-> xs) = %\hspace*{\fill}%-- {\slshape\cstep\label{c:mkMatchR}}
    case x of
      Match _ _ -> mkMatch (f x) xs %\hspace*{\fill}%-- {\slshape\cstep\label{c:nested}}
      _         -> case rhs of
                     [(_,r)] -> r %\hspace*{\fill}%-- {\slshape\cstep\label{c:scalar}}
                     _       -> Case x rhs %\hspace*{\fill}%-- {\slshape\cstep\label{c:case}}
    where
      rhs = [ (trace, mkMatch (f x') xs) %\hspace*{\fill}%-- {\slshape\cstep\label{c:rhs}}
            | trace <- traceR @e         %\hspace*{\fill}%-- {\slshape\cstep\label{c:traceR}}
            , let x' = Match trace x ]   %\hspace*{\fill}%-- {\slshape\cstep\label{c:match}}
\end{haskell}
\caption{Implementation of the \lstinline{match} function, which enables us to
  lift pattern matching from the host language to the embedded language.}
\label{fig:matching}
\end{figure}


Arguably, one limitation of this approach is the need for @match@ to be supplied
a \emph{function} on embedded terms, so that it can be continually reapplied to
explore every success continuation. For example, the user can not write their
embedded program as:
\begin{haskell}
inline%$_1$% x = case f x of
  Nothing_ -> ...    -- error: embedded pattern synonym\ldots
  Just_ y  -> ...    -- \ldots used outside of `match' context
\end{haskell}
Since the @match@ function is not invoked in @inline@$_1$
neither of the pattern matches will succeed. However as embedded pattern synonyms
require their argument to be in the form generated by @match@, we can detect
this usage mistake and present the user with an informative error message. In
order to still make use of inline @case@ statements we rewrite this example as:%
\footnote{Using the syntactic extensions lambda case and block arguments,
together with the reverse application operator \lstinline[style=footnote]{(&) :: a -> (a -> b) -> b}
from the standard module \texttt{Data.Function}.}
\begin{haskell}
inline%$_2$% x = f x & match \case
  Nothing_ -> ...
  Just_ y  -> ...
\end{haskell}


\subsubsection{Nesting \texttt{case} Statements}
\label{sec:nested_case}

As discussed, due to the semantics of pattern matching the @TraceR@
structure necessarily represents an entire pathway through all of the
sub-patterns of a constructor that result in a successful pattern match. Therefore, scrutinising a term of type @Maybe Bool@, for
example, yields a @Case@ term with three success continuations. Similarly, we
can not know if a pattern match succeeded because it matched on the wildcard
pattern `\textunderscore', which can lead to redundant equations in the list of
success continuations.

These limitations are unavoidable during host program generation of the embedded
program. 
Only once the abstract syntax for the embedded program has been generated using
the above procedure can we optimise it by (a) introducing nested case
statements, for example by the method of Wadler~\cite{Wadler:1987_patterns} or
Augustsson~\cite{Augustsson:1985}, and; (b)
introducing default statements to eliminate redundancy, for example by directly
comparing terms for equality.
This is fairly straightforward
and, moreover, orthogonal to the contribution presented in this work, so we
elide the details; the interested reader may refer to our implementation of this
technique in the embedded language Accelerate for a complete example~(\S\ref{sec:evaluation}).

\subsection{Embedded Products, Revisited}

As a final detail, we return to our formulation of embedded pattern
synonyms for product types (\S\ref{sec:products}). Consider:
\begin{haskell}
nested%$_3$% :: %\textcolor{exp}{Exp (Maybe Int, Bool)}% -> %\textcolor{exp}{Exp Int}%
nested%$_3$% x = x & match \case ...
\end{haskell}
When used within a @match@ context (the argument is
wrapped in a @Match@ term), the pair
constructor must propagate the sub-pattern traces to each component of the pair.
Outside of this context the @Match@ term can be ignored.
Similarly, the @traceR@ instance for product types returns the Cartesian product
of the sub-pattern traces.




\section{Case study}
\label{sec:evaluation}
\label{sec:case_study}

We implemented embedded pattern matching in
Accelerate~\cite{Chakravarty:2011,McDonell:2013,McDonell:2015,CliftonEverest:2017,Haak:2020},
an open source language deeply embedded in Haskell for data-parallel array
computations. 
Accelerate
includes a runtime compiler which generates and compiles parallel code using
LLVM~\cite{Lattner:2004}. 
Adding embedded pattern matching to Accelerate allowed us to simplify both the
internals of the compiler, for example by removing @Bool@ as a primitive type,
%
as well as the user facing language. The @stimes@
function from the @Semigroup@ instance changed from:
%
%
\begin{haskell}
stimes n x = lift . Sum $
  fromIntegral n * getSum (unlift x :: Sum (Exp a))
\end{haskell}
to code that, apart from a trailing underscore, is exactly the same as the
definition of this procedure in the host language:
\begin{haskell}
stimes n (Sum_ x) = Sum_ $ fromIntegral n * x
\end{haskell}

While we did not run a controlled user study to systematically evaluate the effect of
these changes on the usability of the language, we do have
anecdotal evidence that it has improved the experience for users.
We have been using Accelerate in a bachelor course for parallel and concurrent
programming with about 200 students. In the first
iteration of the course we used the original version of Accelerate, without
embedded pattern matching, and in
the following years we used Accelerate with our changes implemented. The cohort who used the original
version struggled to understand when and how to use @lift@ and
@unlift@ to work with algebraic data types, and found it difficult to interpret the
error messages that the compiler emitted when these
operations where missing or used incorrectly. These error messages are
difficult to understand (as they expose details of the implementation) and difficult to fix (as
they require knowledge of extensions such as @ScopedTypeVariables@). The later
cohorts experienced none of these issues: once they were told how to use
pattern synonyms---for example to use @T2@ to construct and deconstruct
pairs---they required no further assistance.




For algebraic data types with a single constructor the
code that is generated by the compiler is the same as it was before
our changes. However, for sum data types we are now able to emit
@switch@ instructions directly. This is an example of a situation where the
built-in notion of a @case@ statement makes it possible to generate more
efficient code, for example by producing a lookup table (a single load
instruction) instead of a
sequence of conditional branch instructions. This also
demonstrates why our representation for sum types uses a flattened
structure with a single @TAG@, rather than a nested binary @Either@
type analogously to how product types are represented as nested binary
pairs.

\section{Related Work}
\label{sec:related}


In this paper we used the example of an embedded language constructed in an
\emph{initial} encoding by representing its abstract syntax as data constructors
(\S\ref{sec:expression}). An alternative is to use a
\emph{final} encoding of the abstract syntax as combinator
functions~\cite{Carette:2009}, which can allow more flexibility in
the interpretation of a language built in this style, but does not change what
features of the host language can be captured in that
representation.\footnote{At least for staged/code generating interpretations of
the encoding, which imposes the most restrictions and is the target of this
work.} 
It should therefore be possible to
combine that embedding approach with our technique.

There are several ways to \emph{lift} values from the host language into the
embedded language. 
The approached we demonstrated here also requires some method to \emph{unlift} values, which is
possible for product data types, but not for sum data types. For example,
to access each component of an expression of type @Exp (a, b)@, it would
first be unlifted into a pair of expressions of type @(Exp a, Exp b)@.
Svenningsson~\cite{Svenningsson:2012} always use
this latter form, which avoids the unlifting step but has the disadvantage of duplicating embedded terms and
requires common-subexpression elimination during code generation to recover
shared structures. Both of these approaches have the subtle limitation that the
host language container data type must be polymorphic in \emph{all} of its
arguments.
Our approach
of using pattern synonyms does not have this limitation, and supports
non-polymorphic data types such as the @Point@ type we saw earlier.


Our technique is implemented purely within the concrete syntax of the host
language,\footnote{
While pattern synonyms make the integration seamless they are not
required for the core of the technique. Host languages without pattern synonyms
could integrate this technique via, for example, a macro system.}
but if we are willing to surrender this quality
then other approaches become possible. QDSLs~\cite{Najd:2016} make use of the
quasi-quoting~\cite{Mainland:2007} feature of TemplateHaskell~\cite{Sheard:2002}
to allow the embedded language author to define their own syntax (and associated
parser) for their embedded language, rather than reusing the syntax of the host
language. Quoting is indicated by @[||...||]@ and unquoting by @$$(...)@, which
adds syntactic noise but clearly demarcates between host and embedded code. That
work demonstrates a QDSL-based version of
Svenningsson's~\cite{Svenningsson:2012} embedded language, but unfortunately
does not explore how quotation might overcome that language's limitations.
The 
LINQ framework as used in
\csharp{}~\cite{Meijer:2006} and \fsharp{}~\cite{Syme:2006}, and the Lightweight
Modular Staging (LMS)~\cite{Rompf:2010} framework in Scala, also make use of
quotation techniques.

Haskino~\cite{Grebe:2016} goes further and uses a GHC plugin to
directly manipulate the AST of the host program during Haskell
compilation time.
As with QDSLs, changing or directly manipulating the host
language provides more opportunities for adding embedded language features, but
as we have shown in this work, such drastic measures are not always required.

\section{Conclusion}
\label{sec:conclusion}

We have presented a method for pattern matching on user-defined algebraic data types in embedded languages. This
improves the usability of embedded languages---at last bringing a distinctive and arguably defining feature of
functional programming to embedded languages---and, potentially, improving the quality of code generated by the
embedding. While the syntactic integration between the host and embedded languages depends on features of the host
language---we make extensive use of pattern synonyms provided by languages such as Haskell and Scala, for example---our
key contributions showing how to generate embedded @case@ expressions can be implemented in any general purpose host
language. Although the AST generation has to be side-effect free, as it may be
executed several times, neither
the host nor embedded languages themselves must be side-effect free.



\bibliographystyle{ACM-Reference-Format}

\afterpage{\enlargethispage{-9cm}}
\bibliography{acc-pattern}

\end{document}